\documentclass[aps,prl,twocolumn,groupedaddress,floatfix,showpacs,showkeys,amsmath,amssymb,10pt]{revtex4-1}
\usepackage{graphicx}
\usepackage{bm}
\begin{document}

\title{Design Principles and Physical Properties of Two-Dimensional Heterostructured Borides}

\author{Alejandro Lopez-Bezanilla}
\email[]{alejandrolb@gmail.com}
\affiliation{Theoretical Division, Los Alamos National Laboratory, Los Alamos, New Mexico 87545, USA}
\date{\today}

\begin{abstract}
Principles of design to create dynamically stable transition metal, lanthanide, and actinide based low-dimensional borides are presented.  A charge transfer analysis of donor metal atoms to electron deficient honeycombed B lattices allows to predict complex covalent heterostructures hosting Dirac states. The applicable guidelines are supported with the analysis of phonon spectra computed with first-principles calculations to demonstrate the physical stability of nanometer-thick heterostructures. Similar or dissimilar layered borides can be stacked on top of each other in a layer-by-layer fashion creating an interface that can be fundamentally different from the individual layers, opening a rich playground to explore novel physical properties and new materials. Functionalities such as multiple Dirac states, highly dispersive electronic bands, and decoupled acoustic-optical phonon are studied. The combination of appealing electronic properties and physical realization make of predicted layered borides promising materials to integrate a new generation of two-dimensional materials.
\end{abstract}

\pacs{Valid PACS appear here}
\maketitle


\section{Introduction}

Progress on synthesis routes, ranging from vapor deposition to solution-processing \cite{C6CC09658A}, will allow for expanding the range of devices that build on the advantages of two-dimensional (2D) materials, which will bring enormous benefits for technology industry\cite{C4NR01600A}.
Special efforts are devoted to developing exfoliation and transfer techniques which allow for piling-up several types of layered materials forming heterostructures kept together by van der Waals (vdW) forces. With strong covalent bonding between the atoms within each layer, layered compounds such as NbSe$_2$, MoS$_2$, WTe$_2$, and TaS$_2$, to name a few, share structural similarity but exhibit very different electronic properties, ranging from semiconducting \cite{PhysRevB.8.3719} to metallic \cite{doi:10.1063/1.2407388} behaviour depending on composition, geometry, and thickness. The resulting thick slabs own novel or enhanced properties with electronic structures often distinctive from their layered constituents\cite{Britnell1311,Hamm1298}. 
The broadness in functionality observed on both the monolayers and the heterostructures originates from the quantum confinement of electrons to the 2D geometry, and also from the modulation of the chemical composition and electronic structure at the atomic scale.

Strictly within the flatland, it remains to be explored the role of strong charge transfer (covalent bonding) between stacked layers, as opposed to transient distortions in the distribution of charge density between layers as in vdW heterostructures. 
Solid candidates to form all-covalent 2D materials are borides, whose parent materials exist in bulk form. Borides are widely used as the world's hardest metals, and to store actinide elements in stable compounds\cite{Lupinetti}. Layered borides can be prepared at ambient pressure\cite{C6RA26658D} and, due to their high electrical conductivity, can be easily cleaved and shaped with electric discharge devices, an extended tool in metal manufacturing. Conventional methods for the syntheses of metal borides are based on the high-temperature carbothermal/borothermal reduction of metal oxides with boron-carbon sources such as boron oxide/carbon or boron carbide\cite{QR9662000441}. One might assume intrinsic experimental difficulties to isolate atomic boride monolayers since they cannot slide past each other due to the strong interlayer covalent bonding, but experimental achievements in the exfoliation of metal borides into aqueous dispersion of few-layer-thick structures allows for envisioning purely covalent layered structures \cite{C6RA26658D}.

\begin{figure}[htp]
 \centering
      \includegraphics[width=0.25\textwidth]{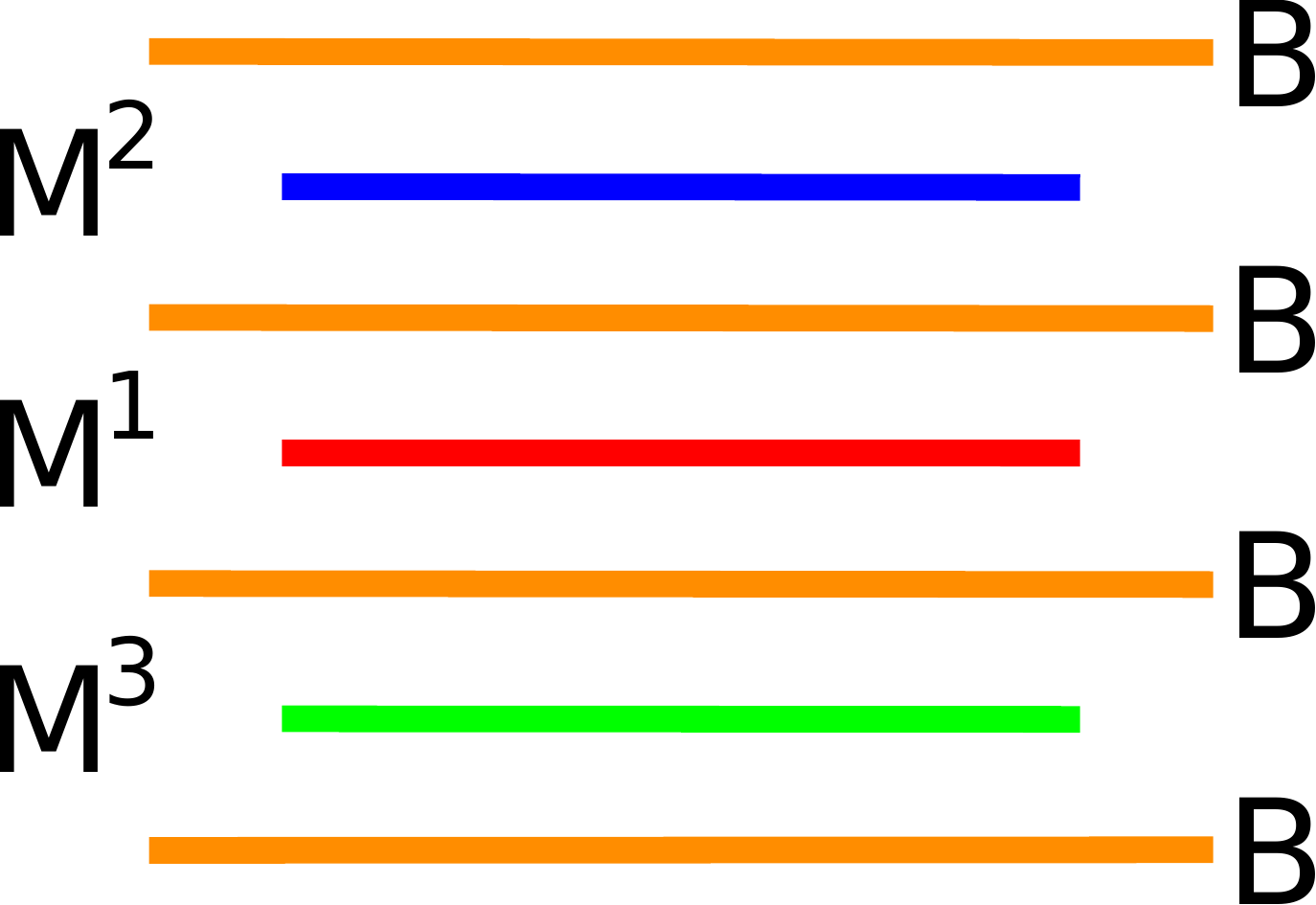}
 \caption{Schematic representation of the multilayered boride systems. Several honeycombed layers of boron atoms (B) sandwich triangular layers of $d$- and $f$-valence orbital based metal elements (M$^n$). }  
 \label{fig1}
\end{figure}

In this paper, we study metal-atom based borides in 2D to analyze their electronic, thermal and dynamical properties, expanding the scope of borides beyond storage and metallurgical purposes.
We branch away from the abundant computational approaches that have propelled the field of materials discovery by seeking optimal configurations relying on little or no empirical knowledge\cite{OganovPickard}. Instead, we propose an insightful analysis driven by an intuition-guided materials informatics scheme to search for structure-property relationships on complex sets of borides. As a result, a broad range of dynamically stable 2D materials hosting multiple cone-shaped bands (Dirac states) are identified. In particular, a principle of design to build $d$- and $f$-orbital-based element borides by piling up layers of existing borides with heterogeneous compositions is procured. A comparison of band and phonon diagrams as well as heat capacity is provided for a series of isoelectronic compounds. The physical viability of the predicted 2D compounds is demonstrated within the harmonic approximation by phonon spectrum analysis.

\section{Computational Methods}
Correlated effects of $f$-orbitals are specially difficult to capture with current first-principles methods, and multiple schemes with different models of screening can be used to provide a description of the band structure of materials based on lanthanide and actinide elements. The approach of choice for this study was self-consistent density functional theory (DFT) with relativistic calculations including spin-orbit coupling (SOC) for all materials. Even though including SOC in a DFT framework may pose some accuracy issues  \cite{PhysRevB.63.035103,doi:10.1080/00018732.2019.1599554}, band structure calculations exhibit expected results such as small band gap openings and mixing of electronic states that remove degeneracies.
DFT-based calculations were performed within the Perdew-Burke-Ernzerhof (PBE) generalized gradient approximation (GGA) functional for the exchange correlation, and the projector-augmented-wave method as implemented in VASP\cite{PhysRevB.48.13115,PhysRevB.54.11169,PhysRevB.59.1758}. The electronic wave functions were computed with plane waves up to a kinetic-energy cutoff of 500 eV. The integration in the 2D k-space was performed using a 56$\times$56$\times$1 $\Gamma$-centered k-point grid. Atomic coordinates and lattice constants were fully relaxed until the residual forces were smaller than 1 meV/\AA. Including an on-site Coulomb repulsion term on the U atom leads to artificial magnetic moments that are absent in experimental observations of for example paramagnetic bulk UB$_2$ \cite{Ohkochi} and, therefore, it was excluded in this study. The force-constant method and the PHONOPY package \cite{phonopy} were employed for computing phonon spectra and heat capacities.

\section{Design principles}
Boron atom is the first and lightest element of group III characterized by an atomic number $Z$=5 and an $s^2p^1$ electronic configuration of its three valence electrons. Boron valence orbitals can combine to form $sp^2$ hybrid orbitals, yielding strong and highly directional $\sigma$ bonds which, together with the short covalent radius and large coordination numbers, allows boron to form a large amount of stable clustered species\cite{BOUSTANI199921}. 
The nature of the bonding in metal borides has been discussed from the point of view of gain or loss of electrons by the B atoms relative to the metal atoms. There is a consensus that electron transfer must be in the same direction for all types of B-rich borides \cite{QR9662000441}, following a simple rule based on the relative electronegativities of B and the d- or f-orbital atoms. 

The so-called 'Aufbau principle'\cite{Boustani} provides construction rules for the design of different highly stable boron structures with quasiplanar geometries that can be viewed as finite-size clusters of planar surfaces. 
The lack of the additional stability provided by the hyper-conjugated network of $p_z$ orbitals typical of C based materials such as graphene or carbon nanotubes is the result of B atom electron deficiency (one valence electron less than C atom) which prevents boron from exhibiting stand-alone honeycombed 2D structures. Although periodic flat distributions including triangular motifs and vacancies\cite{FengB} are possible, realization of 2D extended flat structures composed of B atoms arranged in honeycombed lattices is always subjected to additional electron supply from another source such as a metal substrate\cite{honeycombborophene}.   

The general design rules introduced by Tang and Ismail-Beigi\cite{PhysRevLett.99.115501,PhysRevB.80.134113} clarified that an optimal configuration is obtained if electrons fill all the in-plane $\sigma$ bonding states leaving in-plane antibonding states empty with partial occupancy of the $\pi$ solid-state orbitals.
Self-doping was demonstrated as a possible route to stabilizing 2D and tubular all-B structures.  Detailed explanations were provided\cite{PhysRevLett.99.115501,PhysRevB.80.134113} on the method to obtain 2D metal borides based on a charge transfer model in which a metal atom donates electrons to the boron sublattice, such as layered MgB$_2$. 

Following this metal-boron stabilizing interaction, we demonstrated the viability of 2D ZrB$_2$\cite{zrb2}, CrB$_4$\cite{crb4}, and UB$_4$\cite{Lopez_Bezanilla_2020} and the physical properties derived from the interplay between $p$ and $d$ orbitals.
Continuing along these lines, here we use and extend the charge transfer model of Refs.\cite{PhysRevLett.99.115501,PhysRevB.80.134113,Liu2019} as a general design principle for boron heterostructures with dynamical stability and relevant physical properties based on metal-atom doping. Specifically, we demonstrate with first-principles calculations that $d$-orbital (transition metal) and $f$-orbital (lanthanide and actinide) atoms yield the exact amount of charge to several honeycombed boron lattices so that 2D structures composed of several metal and boron atoms layers are physical viable. This results is consistent with nuclear magnetic‐resonance experiment of the groups IV, V, and VI transition‐metal bulk diborides\cite{doi:10.1063/1.1733774} that suggested that the stability of $\sigma$ and $\pi$ bonds in the boron layer is possible by transfer of electrons in metal-boron donor-acceptor bonds.

In a single B honeycombed lattice with a metal atom (M) sitting on top of each hexagon, transfer of a whole electron per B atom would give a formulation M$^{2+}$(B$^-$)$_2$. B layers can be regarded as isoelectronic with graphene upon charge transfer, which makes Ti, Zr, Hf based monolayers isostructural transition metal borides stable compounds\cite{zrb2}. The resulting material features electronic properties similar to those of graphene such as dispersive Dirac states in the vicinity of the Fermi energy level, with different equivalency and slight mixing of states due to SOC effects.

\begin{figure*}[htp]
 \centering
   \includegraphics[width=0.98\textwidth]{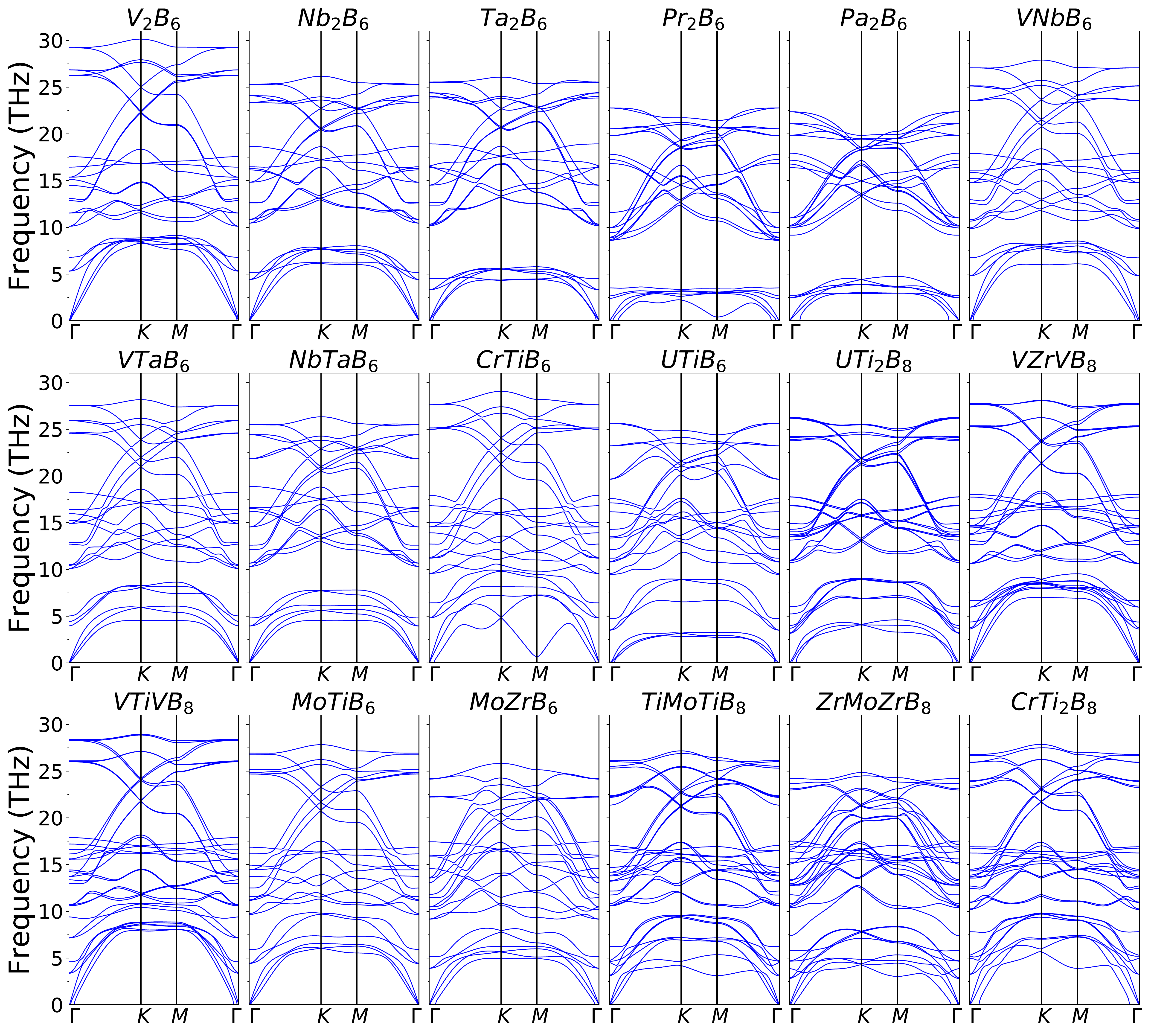}
 \caption{
 Phonon spectra of studied layered borides. The absence of negative frequencies guarantees the dynamical stability of the nanostructures. The same y-range for the single-metal-atom type borides allows to appreciate the effect of an increasing  atomic mass in reducing the phonon dispersion and compacting the total frequency range. In most cases, one or two indirect band gaps are observed between acoustic and optical branches (i.e., UTi$_6$) or optical branches (i.e., NbTaB$_6$). The presence of phonon gaps has a direct consequence on the evolution of the heat capacity with temperature, as shown in Figure \ref{figHeat}.}  
 \label{figPhonons}
\end{figure*}

Deposition of a second hexagonal B layer on top of the triangular lattice of M atoms yields a structure where each M atom is covalently attached to four B atoms simultaneously, two in each layer. The formulation M$^{4+}$(B$^-$)$_4$ entails the utilization of d$^4$ of f$^4$ atoms as stabilizing species of both B lattices. Candidate M atoms are Cr, Mo, W in the d-orbital series, and Nd and U atoms in the f-orbital series. 

Following this metal-boron stabilizing interaction, it is natural to consider complex structures where a M$^{4+}$(B$^-$)$_4$ structure serves as a substrate of additional single layer metal borides. Indeed, over that closed shell compound a M$^{2+}$(B$^-$)$_2$ layer could be deposed, conserving the heterostructure the electronic balance of the parent compounds. According to the representation of Figure\ref{fig1}, the new compound would be represented by M$^1$M$^2$B$_6$. Furthermore, adding another layer to the opposite side of the M$^1$B$_4$ sub-layer would still preserve the charge balance of the new  M$^1$M$^2$M$^3$B$_8$ heterostructure. The theoretical prediction of the viability of these heterostructures does not entail a trivial experimental synthesis but provides evidence of the potential richness of borides when combined to form multi-element 2D compounds. Their dynamical stability is analyzed in the next section to show that despite the disparity of the B and metal-atom masses and radii, the stability is guaranteed as long as the electronic balance is preserved.

\section{Dynamical stability}
Metal borides as well as boron hydrides are some of the few systems in which extensive clustering and catenation of boron atoms occurs\cite{QR9662000441}. In our case, the dynamical stability of the 2D transitional metal borides proposed above is practically straightforward if we consider that some of the single metal-atom compounds analyzed exist in bulk form. For instance, vertical AA stacking of single hexagonal B layers intercalated with Zr or U yields bulk ZrB$2$\cite{Zimmermann20081405} and UB$2$\cite{KARDOULAKI2020153216} respectively, two ceramic materials with a hexagonal covalent structure owning good thermal and electrical conductivities.

\begin{figure}[htp]
 \centering
   \includegraphics[width=0.49\textwidth]{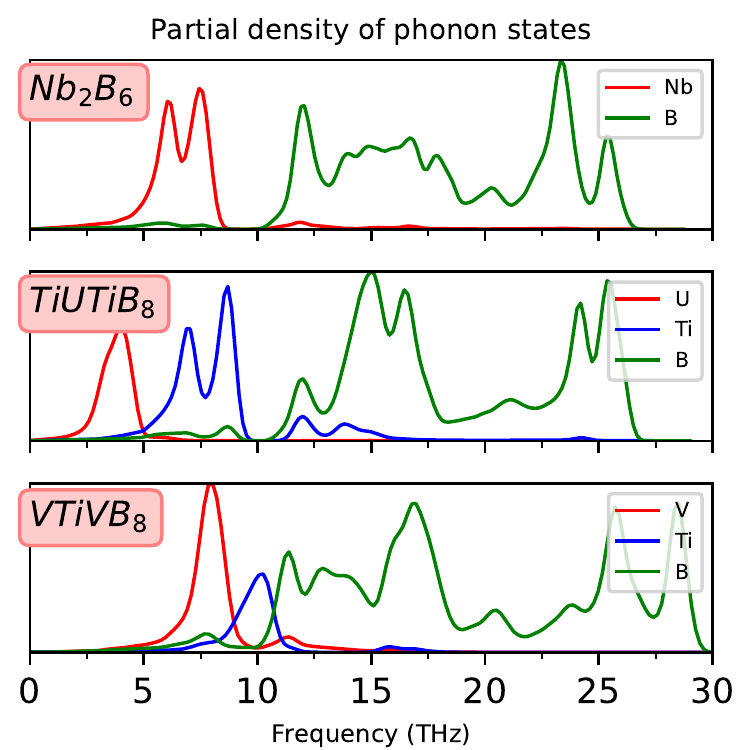}
 \caption{Separate contribution of each atom to the phonon frequencies, represented as projected density of phonon states, of TiUTiB$_8$, Nb$_2$B$_6$, and VTiVB$_8$. Heaviest metal atoms contribute mainly to lowest frequency phonons whereas B atoms dominate the higher frequency modes. y-axis is in arbitrary units.}  
 \label{figPhononspdos}
\end{figure}

Dynamical stability is analyzed within the linear displacement regime at zero temperature for predicting vibration properties. Inspection of phonon diagrams in Figure \ref{figPhonons} allow us to verify that the compounds are stable as the absence of negative vibration frequencies and of anomalous softening modes reveal. Since the analysis of the Zr, Cr, and U single layer borides were presented elsewhere by the author of this paper\cite{zrb2,crb4,ub4}, we will focus hereafter on the new proposed borides. Phonon diagrams of heterostructures are indicated by the notation M$^1$M$^2$B$_6$ and M$^1$M$^2$M$^3$B$_8$, where in some cases M$^1$=M$^2$ (e.g. V$_2$B$_6$) or M$^2$=M$^3$ (e.g. ZrMoZrB$_8$). The Brillouin zone (BZ) of interest with symmetry P6/mm is the area delimited by the $\Gamma$, $M$, and $K$ points in the hexagonal reciprocal lattice.

In all cases, restoring forces oppose deformation when atoms are displaced from their equilibrium positions, indicating that small deformations induced on the material are inefficient in changing their shape. Phonon modes span a frequency range of less than 30 THz, depending the total extension on the mass of the metal atom: the heavier the metal atom(s), the narrower the spectrum extension. Bi- and tri-metallic atom layered structures exhibit phonon diagram with 24 and 33 phonon branches respectively: three acoustic (LA, TA and ZA) and 21 or 30 optical modes. The LA and TA modes correspond to longitudinal and transverse phonon oscillations of the metallic atoms in the 2D layer plane. Mode ZA corresponds to vibrations in the direction normal to the direction of oscillation of the former modes. LA and TA modes progress from a linear dispersion in the zone center ($\Gamma$) to a non-dispersive evolution in the zone edge ($M-K$). Contrary to commonly observed in 2D nanostructures \cite{PhysRevB.84.155413}, layered borides exhibit optical branches of larger dispersion and broader distribution of phonon velocities than acoustical branches. For instance, Ta$_2$B$_6$ exhibits 10 THz dispersive optical modes, whereas the acoustic modes span $\sim$6 THz.
In the single-type metal-atom layers, the heavier the metal atom is, the less dispersive the acoustic phonons are. The $d^3$ and $f^3$ atom series in the first row of Figure \ref{figPhonons} clearly displays the progressive narrowing of the frequency range as one increases the number of protons in the isomorphic structures.

The large difference of masses between B and M atoms determines the predominant contribution of each atom to acoustic and optical phonon branches. Figure \ref{figPhononspdos} shows the separate contribution of each atom to the vibration modes at all frequencies of three borides. The heavier metal atoms contribute mainly to the lowest frequencies, an in Nb$_2$B$_6$, were all phonons branches below 8 THz correspond almost exclusively to vibration modes of the Nb atoms. In UTi$_2$B$_8$ the U atom participates only in the acoustic modes. In turn, Ti atoms are the main participants in the next set of six optical branches. Although B atoms can vibrate at all frequencies, their energetic contribution to the vibration spectrum is mainly above 10 THz. For two metal atoms of similar masses, the difference is less pronounced but significant, as in VTiVB$_8$, where both V and Ti peaks in the projected density of states differ by 2.3 THz. Visual inspection of the eigenvectors\cite{henriquemiranda} generated by PHONOPY reveals that at lower frequencies metal atoms vibrate with larger amplitude than B atoms. Flat branches right above the acoustics correspond to vertical oscillations of M atoms within the honeycomb layers, which explains the low propagation character of those branches. At higher frequencies B atoms vibrate with a remarkably large amplitude while M atoms are barely displaced. 

The large difference of the atomic masses is the origin of an indirect gap between optical and acoustic modes in some structures, similarly to the gap observed in bulk BAs\cite{PhysRevLett.109.095901,PhysRevLett.111.025901,Tian582} and Li$_2$Te\cite{parker}.  This gap and that gap found in the rest of the structures between bunched optical branches enables selective phonon frequency transmission, namely no kind of atomic vibration propagation is allowed in the band gap. In particular UTiB$_6$ displays a tiny gap (which vanishes when adding a second layer of TiB$_2$) that would the acoustic modes avoid crossing with optical modes. This may potentially lead to interesting phenomena in thermal transport, since acoustic-optical phonon backscattering is not allowed by the gap\cite{PhysRevLett.109.095901}.

\begin{figure}[htp]
\centering
\includegraphics[width=0.49\textwidth]{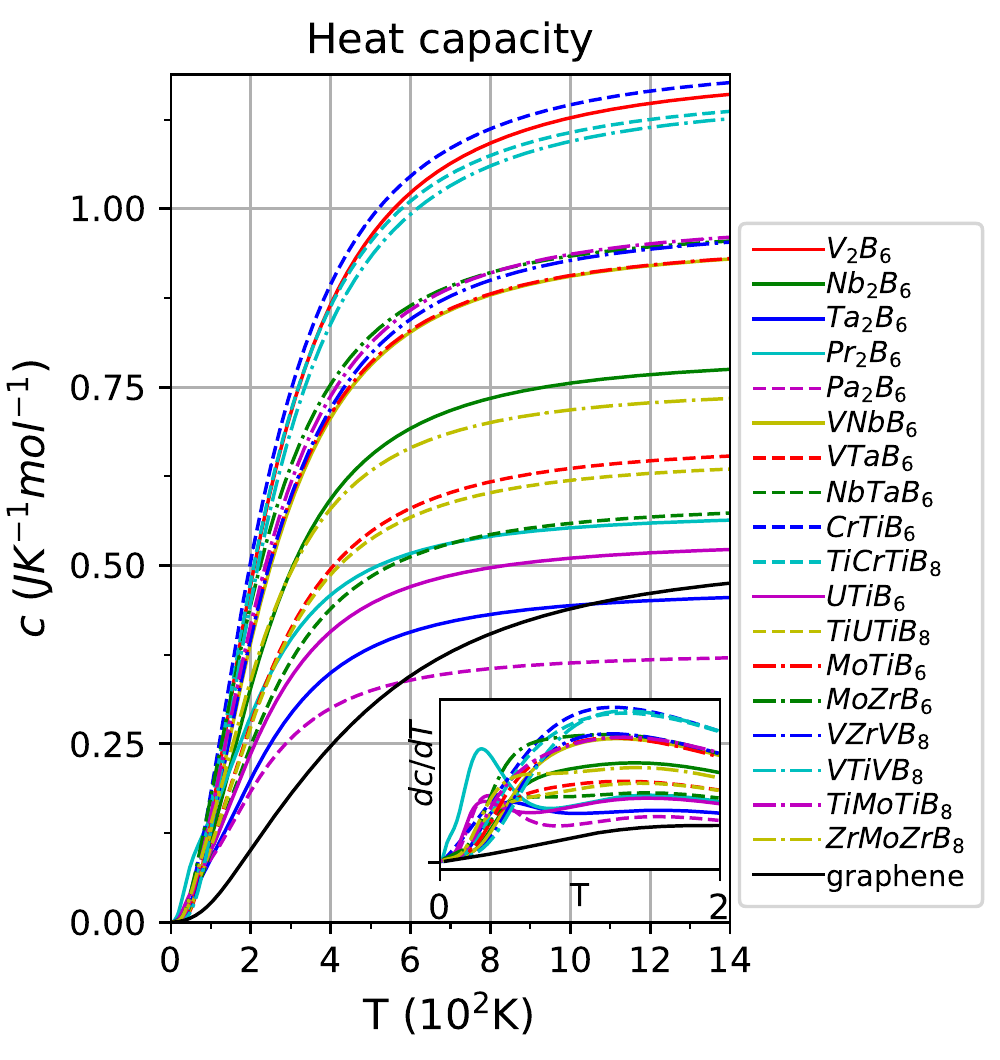}
     \caption{Specific heat capacities $c$ of the multilayered B compounds normalized to the mass in each unit cell are compared with the one of graphene. Inset: The derivative of $c$ with respect to the temperature shows that $c$ increases differently at several T ranges as a consequence of the various gaps observed in the phonon spectra. }
\label{figHeat}
\end{figure}

Figure \ref{figHeat} shows the dependency of the specific heat capacity $c$, in $J K^{-1} mol^{-1}$ and normalized to the atomic weight $m$ of each cell, with temperature of all the borides. Specific heat capacity is defined as the ratio of the supplied heat to the resulting temperature change for a unit mass of the substance and, therefore it measures the ability of the materials to absorb heat. Dimensional restriction, small group velocity, and large mass of the metallic atoms lead to high specific heat capacities. Lighter metal based borides exhibit enhanced $c$ at any temperature with respect to the heavier-metal compounds, exceeding in all cases that of graphene below 600 K, also shown for comparison in Figure \ref{figHeat}. Each bundled curves shows how much of an effect metal atoms can have on the heat capacity of the borides, as the main difference at temperatures above $\sim$ 100 K is in the composition. 
At $\sim$100 K the normalized $c$ of all borides are very similar and much larger than that of graphene, whereas at $\sim$ 200 K the $c/m$ of lighter compounds starts diverging. At large temperature, the specific heat capacities serves to differentiate the mass composition of the layered borides. 

At very low temperatures, the heat capacity of some borides undergo two slope changes. For a clearer representation, in the inset of Figure \ref{figHeat} the derivative of $c/m$ with respect to the temperature is plotted. With two local maxima, first at $\sim$33 K and second at $\sim$130 K, this anomalous change is due to the frequency gap between phonon branches that can also be observed for compounds such as Pr$_2$B$_6$ and UTiB$_6$, and is absent in the graphene layer.

\section{Electronic properties}

\begin{figure*}[htp]
 \centering
\includegraphics[width=0.98\textwidth]{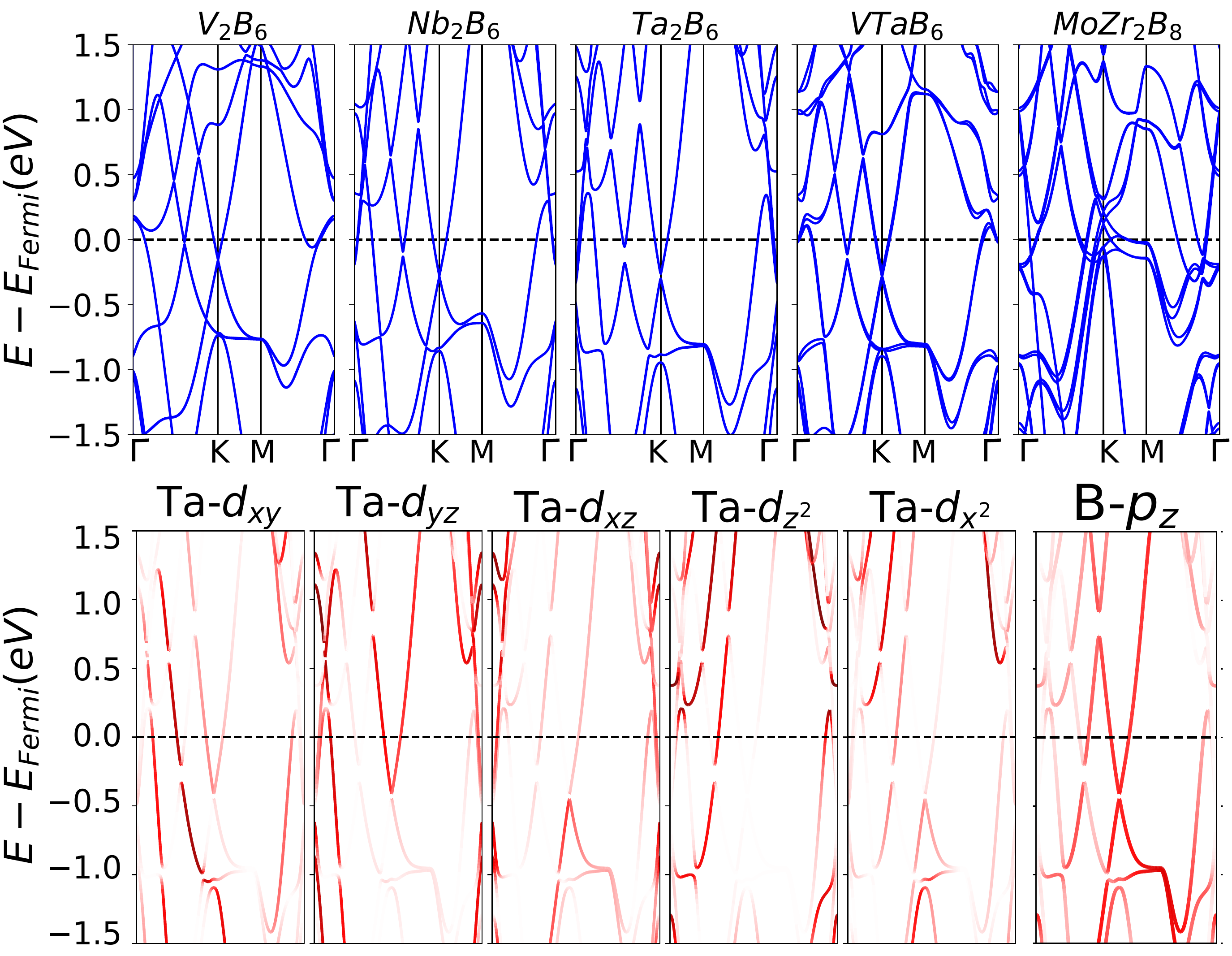}
 \caption{Electronic band diagram of  V$_2$B$_6$, Nb$_2$B$_6$, Ta$_2$B$_6$, VTaB$_6$, MoZr$_2$B$_8$. Two Dirac cones in the vicinity of the Fermi level develop a gap as the spin-orbit coupling effects become larger with metallic atoms size. Very dissimilar (VTaB$_6$) and similar atomic radius (MoZr$_2$B$_8$) lead to a very different distribution of electronic bands despite the charge transfer model is the same for all the borides. Color-resolved electronic state projection (lower panels) of Ta$_2$B$_6$ show the participation of each atomic orbital to the material's electronic states. The color intensity of the lines is proportional to the contribution. Horizontal dashed lines point out the Fermi level.}  
 \label{figBands}
\end{figure*}

Low-energy fermionic excitations behaving as massless Dirac particles is a subject of intense study for multiple technological applications ranging from quantum information technologies to novel electronic devices. Graphene\cite{Novoselov666} is one the prototypical materials hosting electronic bands with linear dispersion in which charge carriers travel at the speed of light due to their zero mass\cite{NovoselovGeim2005}. The electronic properties of the heterostuctured layered borides studied in this paper resemble to monolayer graphene in the shape of their electronic bands and the presence of nodes in the excitation spectrum as a result of groups of two cone-shaped bands touching each other at one single point in the BZ. A relevant difference of the borides with respect to graphene is the number of such cones and the location in the BZ. Whereas graphene exhibits two Dirac points at K and K' points, borides outnumber with some electronic bands out of high-symmetry points in the hexagonal BZ. The field of 2D Dirac materials owning cone-shaped band structure and high carrier mobilities is expanded here with the description of multiple borides that with strong or weak SOC exhibit similar features to that of graphene. 

Figure \ref{figBands} shows the relativistic band structure of a selection of the borides. Focus is on the lighter metal atom borides with no $f$-orbitals, which makes DFT calculations less controversial. Indeed, multiple schemes with different models of screening can be used to provide a first-principles description of the band structure of correlated $f$-orbitals rich borides, they all leading to various interpretations of results that as of today cannot be compared with experimental evidence on thin metal borides. Relativistic effects due to the large size of metallic elements are accounted in the calculations including SOC effects.

First three panels of Figure \ref{figBands} show the distribution of electronic states of the group-V elements with $d^3$ electrons. According to our charge transfer model, a total of six $d$-electrons are yielded to the six empty $p_z$ orbitals of the B unit-cell atoms in the three honeycombed lattices. As opposed to graphene, several bands cross the Fermi level. Similarly to the all-C monolayer a Dirac point is observed at K point, although slightly below the Fermi level. In addition, a second set of cone-shaped bands meet at a single point at Fermi level in the $\Gamma - M$ line. This picture is clearly visible in V$_2$B$_6$. As the mass of the metal atom increases, a small gap between Dirac cones develops, becoming a few meV large in Nb$_2$B$_6$, and clearly noticeable in the band diagram of Ta$_2$B$_6$. The combination of two metal atoms with dissimilar size and a relative lack of symmetry in the material leads to a splitting of electronic bands and removal of state degeneracy, as shown in VTaB$_6$. However, a heterostructure composed of two similar-size atoms in a more symmetric geometry such as MoZr$_2$B$_8$, partially preserve the dispersion of the bands found in the one-metal-atom type borides. 
 
It is worth determining the origin of the Dirac cones, and Ta$_2$B$_6$ is taken as a case of study. Lower panels of Figure \ref{figBands} display a color-scheme representation of the electronic state diagrams projected onto each band, which allows for observing the independent contribution of each M-$d$ and B-$p_z$ atomic orbital. At K point, the main contribution to the upper cone-shaped bands is from Ta-$d_{yz}$ orbitals, whereas Ta-$d_{xz}$ orbitals are involved in the lower cone. Two different Ta orbitals are responsible of the two cones in the middle of the high-symmetry $\Gamma - M$ line, namely Ta-$d_{xy}$ and Ta-$d_{z^2}$. Note that whereas the cones at K point can be projected on two cone-shaped contributions, the second two cones are projected in a state that develops from near the zone center in the conduction bands down to the vicinity of K (Ta-$d_{xy}$), and a state that extend from the valence band up to high energies, with two gaps (Ta-$d_{z^2}$) in the energy range displayed. From the B atom, the $p_z$ orbital also contribute to the delocalization of charge projecting its contribution to the four cones, pointing out to strong hybridization of this orbital with all Ta-atom orbitals.

\section{Conclusions}
A general strategy to mix and match layered boride materials with different properties has been presented. Covalent heterostructures with a variety of dynamical and electronic properties have been proposed benefiting of the ability of 2D layer of similar and dissimilar composition to attached each other.  Theoretical evidence has been provided on the following: i) a series of borides constructed following a charge transfer model from metal to B atoms exhibit dynamical stability as 2D compounds, ii) an itinerant character of the charge carriers is observed due to the B honeycombed layers are isolectronic to graphene, hosting electronic states with a delocalized character, iii) the heat capacity of the thin layers is the largest observed in 2D materials. iv) borides feature an infrequent acoustic-optical and optical-optical gap as a result of the large difference in the mass between the metal and B atoms.

The large variety of compounds in which layered borides can take form leads to a rich diversity of physical phenomena usually not found in 2D materials, such as multiple Dirac states and bundling of acoustic and optical phonon branches. Although the most accurate predictions must come upon contrast with experimental observation, first-principles calculations support our charge transfer model, providing a theoretical ground to predict new compounds. The design principles of this study contribute to the creation of a theoretical framework to predict and explain the physics of borides in 2D, where $d$ and $f$-orbitals have a prominent role in the formation of Dirac states. The predicted outstanding electronic properties of heterostructured borides provide an extra motivation for borides integration with the  nanometer-scale silicon complementary metal-oxide-semiconductor (CMOS) technology as well as beyond-CMOS devices and circuits. 
\cite{Novel2DMB2}

\section{Acknowledgments}
Los Alamos National Laboratory is managed by Triad National Security, LLC, for the National Nuclear Security Administration of the U.S. Department of Energy under Contract No. 89233218CNA000001. 
I acknowledge the computing resources provided on Bebop, the high-performance computing clusters operated by the Laboratory Computing Resource Center at Argonne National Laboratory.

\end{document}